# KubeEdge.AI: AI Platform for Edge Devices

Sean Wang
Cloud Strategic Planning
Futurewei Technologies, Inc.
Bellevue, WA, USA
swang1@futurewei.com

Yuxiao Hu
Cloud Strategic Planning
Futurewei Technologies, Inc.
Bellevue, WA, USA
yuxiao.hu1@futurewei.com

Jason Wu
Cloud Strategic Planning
Futurewei Technologies, Inc.
Bellevue, WA, USA
jason.wu@futurewei.com

## ABSTRACT

The demand for smartness in embedded systems has been mounting up drastically in the past few years. Embedded system today must address the fundamental challenges introduced by cloud computing and artificial intelligence.

KubeEdge [1] is an edge computing framework build on top of Kubernetes [2]. It provides compute resource management, deployment, runtime and operation capabilities on geo-located edge computing resources, from the cloud, which is a natural fit for embedded systems.

Here we propose KubeEdge.AI, an edge AI framework on top of KubeEdge. It provides a set of key modules and interfaces: a data handling and processing engine, a concise AI runtime, a decision engine, and a distributed data query interface. KubeEdge.AI will help reduce the burdens for developing specific edge/embedded AI systems and promote edge-cloud coordination and synergy.

## 1 Challenges

KubeEdge.AI addresses following challenges for embedded AI systems:

**Data is essential:** the importance of data is emphasized in every aspect. First, it is collected in much larger volume with greater focus on non-structured data, which poses challenges to both storing and processing. Secondly the usage of such data has much wider variety. Not only locally, geo-located data could have greater value when cross referenced and queried. Last but not the least, data security and privacy are closer more than ever to our real life, which need to be protected at all cost.

**Cognitive capability:** the need to understand our physical surrounding and interact with it is a key ability for the goal of digital transformation of the world around us, embedded system naturally stands on the frontline of this challenge.

**Decision making capability:** to solve real world problems, AI more than ever needs to be brought to where real events happen, building artificial intelligence system with decision making capability is essential for the goal.

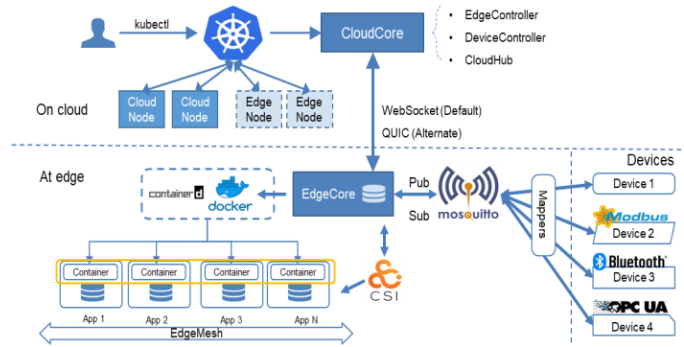

**Figure 1 KubeEdge Architecture**

**Edge-Cloud Synergy:** imagine the difficulty of managing and utilizing millions of compute modules and billions of devices. The name edge is defined in related to the cloud, which can be terminals, mobile phones, or any devices with embedded computing capabilities. The demand of compute power at edge does not diminish the importance. Best result can only be achieved by having both edge and cloud working together. Cloud provides the central management location and supports edge components with its powerful and elastic computing power.

## 2 Major Components of KubeEdge.AI

The architecture of KubeEdge and the major components of KubeEdge.AI are shown in **Figure1** and **Figure2**.

## 2.1 Data Engine

In embedded AI systems, especially in industry scenarios, data are collected on multiple sensors. These time sensitive data need to be processed in real time for logging, monitoring, alerting and automations. Furthermore, for future offline investigation and global analysis, these sensor data need to be stored locally and uploaded to central storage, with sampling, alignment, encryption, and other necessary preprocessing steps. These are the main differentiators comparing to tradition IoT system. A time-series database (TSDB) at edge provides a space efficient engine to store, process and query real time data during above process. Besides, a computing engine for streaming data and/or interactive/batch data processing will also be involved. So, the proposed KuberEdge.AI platform integrate TSDB, Flink and Spark to handle these requirements.

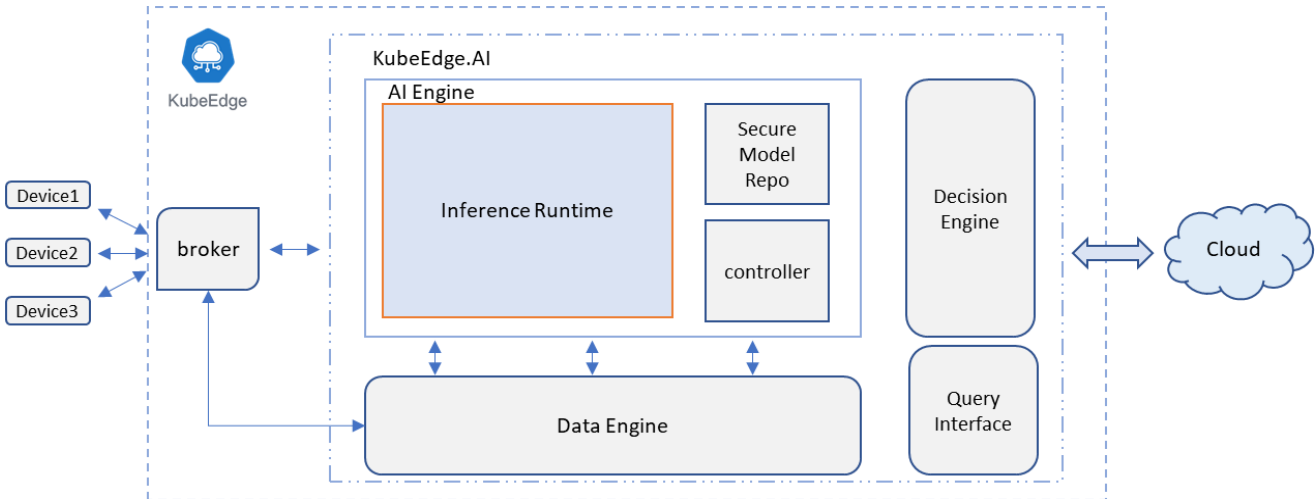

**Figure 2. KubeEdge.AI Framework**

## 2.2 AI Engine

To enable intelligence on an embedded system, each node will be equipped with AI processing capabilities, including both hardware and software. Recently, variety of AI chips have been released, such as Ascend by Huawei, Sunrise by Horizon, QuestCore by Yitu and HanGuang by Ali etc.[3]. Machine learning software frameworks such as SparkML, TensorFlow and MindSpore [4] are also developed and open sourced to support training and inference for classical and latest AI algorithms. Limited by power consumption, form factor, and deployment locations, embedded AI systems require certain functionalities different from general AI systems running on public cloud, or private data centers. KubeEdge.AI addresses them by leveraging following modules:

*2.2.1 Model Deployment.* AI algorithms will be trained on public or private cloud, adapted according to target scenarios and packaged as models. The models can be classical machine learning algorithms such as logistic regression and SVM, or latest deep learning algorithms such as MobileNets and Mask R-CNN. KubeEdge.AI provides mechanics to deploy them to multiple edge devices on demand, according to the data source type and query definition.

*2.2.2 Model Refreshment.* Models need to be updated to make embedded AI system agile and data-driven. KubeEdge.AI uses its AI engine to track model versions and provide different model update schema, including immediate update, lazy update and scheduled update, etc.

*2.2.3 Inference Runtime.* After an AI model is deployed to embedded nodes, inference happens when the specified source data arrives. The result will be stored in TSDB on the node and uploaded to central storage as scheduled or on demand.

*2.2.4 Privacy and Security.* In most industrial scenarios, data collection and model training are conducted by different parties and at different locations. Inspired by federated learning, we implement protocols in KubeEdge.AI to ensure both the model security and data privacy are protected.

## 2.3 Decision Engine

Besides monitoring and diagnostic, another important goal of an embedded AI system is to help users make decision intelligently and autonomously, e.g. to predict the future production rate, to decide the best system configurations, etc. KubeEdge.AI platform integrates a decision engine which supports both rule-based and data-driven methods.

## 2.4 Query Interface

Edge node cannot be isolated, nodes in group by vicinity or other attributes could be required to work together in order to achieve certain tasks beyond single node boundary. Moreover, geo-scattered data gives different challenges to analytics on higher and aggregated level. KubeEdge.AI provides a SQL-like query interface to extract data and performance local queries.

## 3   Current Status and Future Works

KubeEdge.AI can be applied to wide range of embedded applications. The commercial product rooted KubeEdge has been used for smart cameras, highway flow control system and more. KubeEdge is an ongoing Cloud Native Computing Foundation (CNCF) project. It is currently running on Linux, with a small footprint and flexible yet powerful computing and communication capabilities. We are focusing on building the foundation components of KubeEdge.AI, including TSDB, processing engine and basic AI capability etc. Full AI engine and end-to-end flow will be built in next steps.

## REFERENCES
[1]  https://kubeedge.io/en/
[2]  https://kubernetes.io/
[3]  https://equalocean.com/ai/20190514-dozens-of-ai-companies-launch-ai-chips-to-win-in-the-future
[4]  Michael J. Garbade, Top 8 open source AI technologies in machine learning